\documentclass[aps,reprint]{revtex4-1}

\usepackage{commath}
\usepackage{amsmath}
\usepackage{graphicx}
\usepackage{hyperref}
\usepackage{txfonts}

\draft 

\begin{document}


\title{A universal self-amplification channel for surface plasma waves}


\author{Hai-Yao Deng$^{1,2}$}
\email{h.deng@exeter.ac.uk}
\author{Katsunori Wakabayashi$^{1,3}$}
\email{waka@kwansei.ac.jp}
\author{Chi-Hang Lam$^4$}
\email{c.h.lam@polyu.edu.hk}
\affiliation{$^1$Department of Nanotechnology for Sustainable Energy,
School of Science and Technology, Kwansei Gakuin University, Gakuen 2-1,
Sanda 669-1337, Japan}
\affiliation{$^2$Department of Physics and Astronomy, University of Exeter, Stoker Road EX4 4QL Exeter, United Kingdom}
\affiliation{$^3$National Institute for Materials Science (NIMS), Namiki 1-1, Tsukuba 305-0044, Japan}
\affiliation{$^4$Department of Applied Physics, The Hong Kong
Polytechnic University, Hung Hum, Hong Kong}  


\begin{abstract} 
We present a theory of surface plasma waves (SPWs) in metals with arbitrary electronic collision rate $\tau^{-1}$. We show that there exists a universal intrinsic amplification channel for these waves, subsequent to the interplay between ballistic motions and the metal surface. We evaluate the corresponding amplification rate $\gamma_{0}$, which turns out to be a sizable fraction of the SPW frequency $\omega_s$. We also find that the value of $\omega_s$ depends on surface scattering properties, in contrast with the conventional theory. 
\end{abstract}


\maketitle 

\section{introduction}
\label{sec:1}
Collective electronic oscillations on the surface of metals,
dubbed surface plasma waves (SPWs)~\cite{ritchie1957,ferrell1958,raether1988}, have emerged as a pivotal player in \textit{nano}scopic manipulation of light~\cite{benno1988,zayats2005,maier2007,ozbay2006,mark2007}. The functionality of many prototypical nanophotonic devices critically relies on the distance SPWs can travel before they are damped out due to energy losses via several channels~\cite{ozbay2006,mark2007,barnes2003,barnes2006,ebbesen2008,martino2012}. SPWs can lose energy due to Joule heat, inter-band absorption, radiation emission and individual electronic motions (Landau damping). Most of the losses can be efficiently but not totally reduced under appropriate circumstances. Amplifiers have been contrived to compensate for the losses~\cite{bergman2003,seidel2005,leon2008,leon2010,yu2011,pierre2012,fedyanin2012,cohen2013,aryal2015}, which are all extrinsic and require external agents such as a dipolar gain medium to supply the energy. 

The standard theory of SPWs was formulated shortly after their discovery in 1957~\cite{ritchie1957} and has been comprehensively discoursed in many textbooks~\cite{raether1988,zayats2005,maier2007,pitarke2007,sarid2010}. In this theory, the electrical properties of metals are effected by a frequency-dependent dielectric function $\epsilon$. To analytically treat $\epsilon$, the simple Drude model or the slightly more involved hydrodynamic model~\cite{pitarke2007,harris1971,fetter,schnitzer2016} is invariably invoked. For either model to be valid, electronic collisions must be sufficiently frequent so that the electronic mean free path, $l_0=v_F\tau$, where $v_F$ is the Fermi velocity and $\tau$ the relaxation time, is much shorter than the SPW wavelength or the typical length scale of the system. The general case with arbitrary $\tau$, especially the collision-less limit, where $\tau\rightarrow\infty$, defies these models and has yet to be entertained. Other models based on \textit{ab initio} quantum mechanical computations~\cite{pitarke2007,peter1984} are helpful in understanding the complexity of real materials but falls short in providing an intuitive and systematic picture of SPWs underpinned by electrons experiencing less frequent collisions. 

In the present work we employ Boltzmann's transport equation to derive a theory of SPWs in metals with arbitrary $\tau$. Our analysis reveals a universal intrinsic amplification channel for SPWs. The existence of this channel does not depend on $\tau$ but is warranted by a general principle. We show that the unique interplay between ballistic electronic motions and boundaries results in an electrical current that allows a net amount of energy to be drawn from the electrons by SPWs, which would inevitably self-amplify in the collision-less limit, thereby destabilizing the system. The hereby predicted self-amplification of collision-less SPWs is analogous to the Landau damping in collisionless bulk plasma waves~\cite{landau1946,wong1964,bingham1997,natu2013}. While the latter is caused by slowly moving electrons that strip energy from the wave, the former is attributable to ballistically moving electrons imparting energy to the wave when boundaries are present. 

In the next section, we describe the system under consideration, state our main results and conceive their possible experimental signature. In Sec.~\ref{sec:theory}, the SPW theory is presented. In Sec.~\ref{sec:3}, we detail the methods used in the theory. We discuss the results and conclude the paper in Sec.~\ref{sec:4}. Some calculations and arguments not covered in Sec.~\ref{sec:theory} and \ref{sec:3} are given in appendices~\ref{sec:edf} and~\ref{sec:M}. 

\section{results}
\label{sec:2}
\textit{System} - We consider a prototypical system, namely, a semi-infinite metal (SIM) occupying the half space $z\geq 0$ and interfacing with the vacuum at a geometrical surface $z=0$. A SIM is not sheerly of academic interest: it can be regarded as one half of a thick metal film. In the spirit of the so-called jellium model,~\cite{ziman1960,pines,abrikosov} the metal is treated as a free electron gas embedded in a static background of uniformly distributed positive charges. On the whole the system is neutral. The kinetic energy of the electrons is written $\varepsilon(\mathbf{v})=\frac{1}{2}m\mathbf{v}^2$, where $m$ and $\mathbf{v}$ denote the mass and velocity of the electrons, respectively. Inter-band transitions are accordingly ignored but their effects will be discussed in Sec.~\ref{sec:4}. The surface is of a hard-wall type and prevents electrons from leaking out of the system. Throughout we reserve $\mathbf{r}=(x,y)$ for planar coordinates while $\mathbf{x} = (\mathbf{r},z)$ denotes the complete position vector. We neglect retardation effects in total. 

\begin{figure*}
\begin{center}
\includegraphics[width=0.97\textwidth]{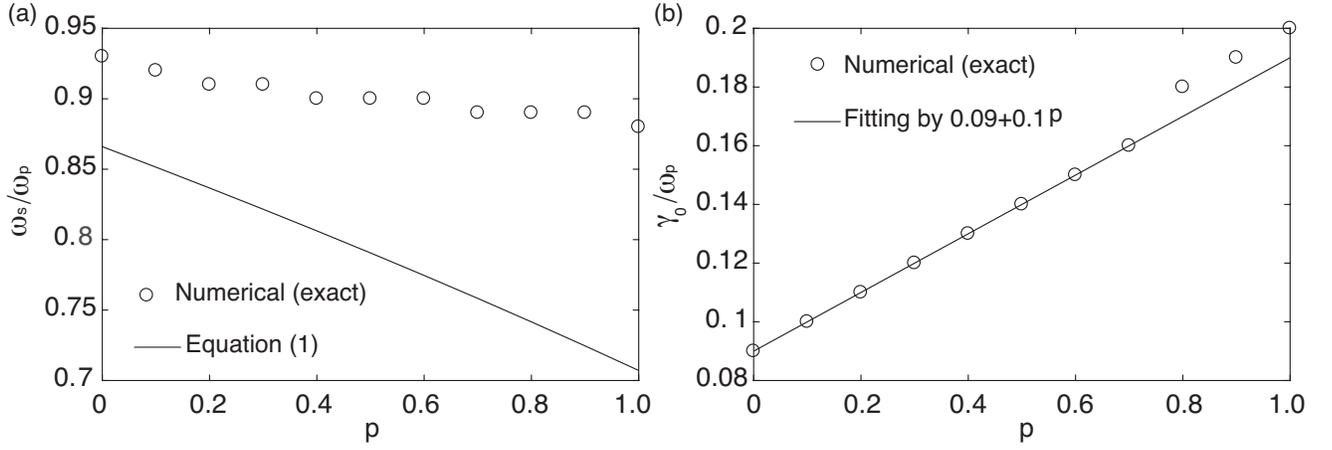}
\end{center}
\caption{Numerically calculated SPW frequency $\omega_s$ and its intrinsic amplification rate $\gamma_0$ as a function of the \textit{Fuchs} parameter $p$, by equation (\ref{spw}). $k/k_s=0.1$ and $q_c/k_s=1.5$. The approximate analytical solutions are also shown. \label{figure:f0}}
\end{figure*}

\textit{Key results} - We find that the SPW frequency $\omega_s$ and its amplification rate $\gamma$ can be written in the following form, 
\begin{equation}
\omega_s = \omega_{s0}~\beta(p), \quad \gamma = \gamma_0 -\frac{1}{\tau}, \quad \gamma_0>0. \label{key}
\end{equation}
Here $\omega_{s0} = \omega_p/\sqrt{2}$ is the SPW frequency in the hydrodynamic/Drude model, with $\omega_p$ being the characteristic frequency of the metal, $p\in[0,1]$ is a parameter that accounts for surface scattering effects, $\beta(p) \sim 1$ weakly depends on $p$ and $\gamma_0$ includes Landau damping and is exactly independent of $\tau$. Both $\omega_s$ and $\gamma_0$ are determined by the secular equation~(\ref{spw}) given in Sec.~\ref{sec:3}. Analytical expressions can be found for $\beta(p)$ and $\gamma_0$ under certain approximations. Exact $\omega_s$ and $\gamma_0$ have been computed numerically and are displayed in Fig.~\ref{figure:f0}, where we observe that $\gamma_0/\omega_{s0}\approx \alpha(1+p)$, with $\alpha \approx 0.1$.  

Equation (\ref{key}) furnishes an intrinsic amplification channel $\gamma_0$ for SPWs. It also shows that, in contrast to the hydrodynamic/Drude model, $\omega_s$ relies on surface properties via the parameter $p$. In particular, as seen in Fig.~\ref{figure:f0} (a), $\omega_s$ is significantly -- more than $10\%$ -- larger than $\omega_{s0}$, the value expected from the hydrodynamic/Drude model. In other words, $\beta(p)$ is unity in these models whereas it could reach up to $1.2$ in our theory. Such a great contrast would be ideal for experimentally verifying the theory. Unfortunately, in common materials such as noble metals, due to pronounced inter-band transitions, there is no simple relation between $\omega_s$ and $\omega_p$. 

\begin{figure}
\begin{center}
\includegraphics[width=0.45\textwidth]{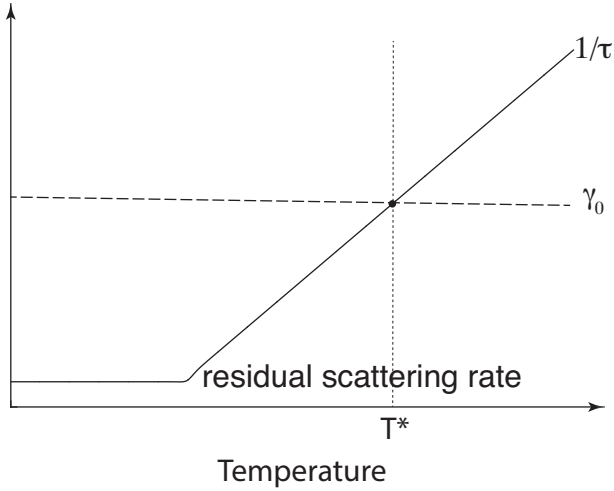}
\end{center}
\caption{Possible instability caused by the self-amplification channel upon cooling down the system. At the critical temperature $T^*$, $\gamma_0 = 1/\tau$. Here the sketch assumes that $1/\tau$ is dominated by phonon and impurity scattering. \label{figure:f3}}
\end{figure}

\textit{Experimental signature} - We conceive one possible repercussion of the self-amplification channel, noting that the net amplification rate $\gamma = \gamma_0-1/\tau$ and its temperature dependence can be directly measured in various ways, e.g. by examining the SPW propagation distance or the width of the energy loss peaks in electron energy loss spectroscopy (EELS) or the quality of the reflectance dips in the Kretschmann-Otto configuration. In this paper, our calculation of $\gamma_0$ is done at zero temperature. However, arguably $\gamma_0$ could bear a different -- probably weak -- temperature dependence than $1/\tau$. In sufficiently pure samples, in which the residual scattering is small enough, there might exist a critical temperature $T^*$, above which $\gamma<0$ while below it $\gamma>0$, i.e.  $T^*$ marks a transition of the system from the Fermi sea to another state, see Sec.~\ref{sec:4} for further discussions. In Fig.~\ref{figure:f3}, we sketch the situation described here, assuming $1/\tau$ is dominated by phonon and impurity scattering.  

To experimentally investigate the self-amplification channel using the most experimented materials like gold and silver, we must clarify the effects of inter-band transitions, which not only bring about additional energy losses but also strongly affect the value of $\omega_s$. We discuss these effects further in Sec.~\ref{sec:4}.

\section{Theory}
\label{sec:theory}
In this section, we give a systematic exposition of the theory that supports equation~(\ref{key}). Some technical aspects are left to Sec.~\ref{sec:3}. The overall objective is to set up the equation of motion for the charge density, analyze its internal structure and solve it. We first discuss the relation between the charge density and the electric field, and then prescribes the current density and show that it possesses a spectacular structure. We proceed thence to the equation of motion, extracting its solutions and unveiling the properties of SPWs. 

\begin{figure*}
\begin{center}
\includegraphics[width=0.95\textwidth]{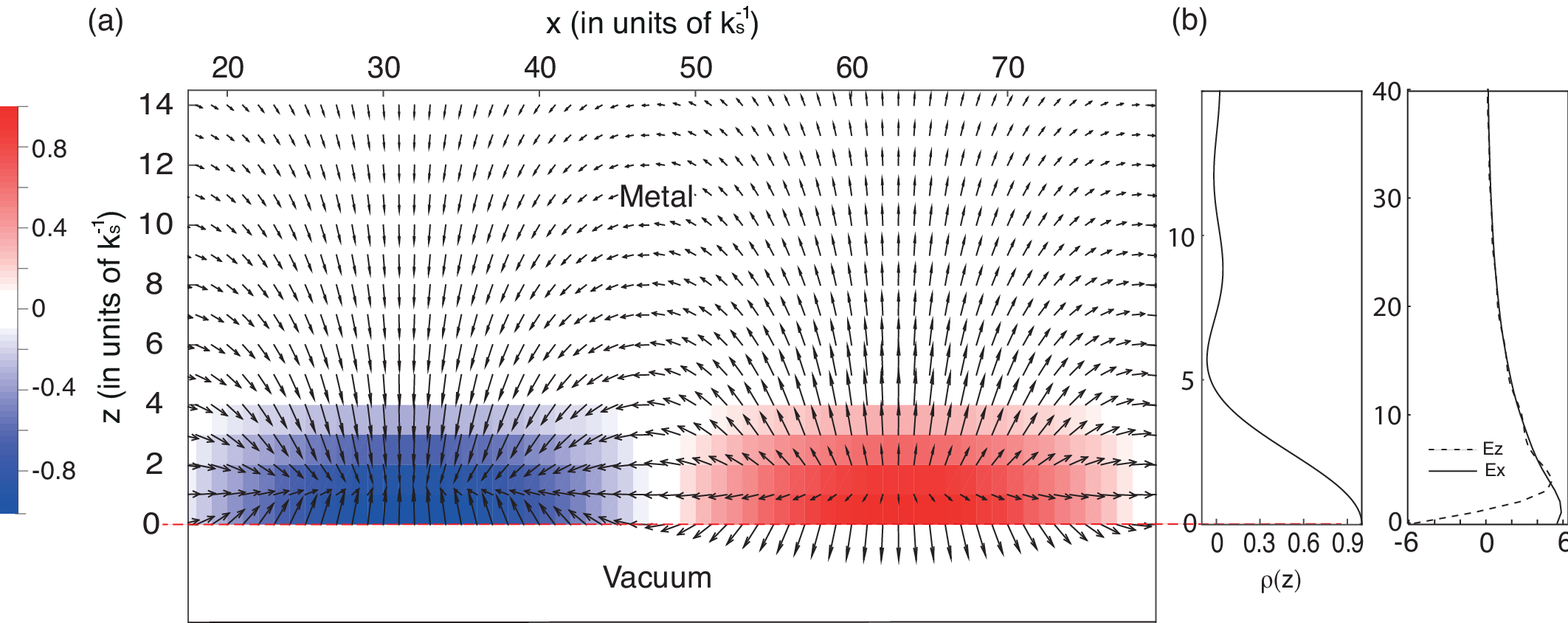}
\end{center}
\caption{Sketch of SPWs supported on the surface of a semi-infinite metal. $k/k_s=0.1$ and $p=1$. (a) Charge density map (color) and electric field map (arrows). (b) Plots of $\rho(z)$ and $E_{\mu}(z)$, where $\mu=x,z$. $\rho(z)$ is calculated by Eq.~(\ref{rho}) while $E_{\mu}(z)$ by (\ref{ex}) and (\ref{ez}). Both $\rho(z)$ and $E_{\mu}(z)$ have been normalized. \label{figure:f1}}
\end{figure*}

\textit{Electrostatics} - We aim for establishing the equation of motion for the charge density $\rho(\mathbf{x},t)$ under the influence of its own electric field $\mathbf{E}(\mathbf{x},t)$. Here $t$ denotes the time. Thanks to the linearity and symmetries of the system, we can write $\rho(\mathbf{x},t) = \mbox{Re}\left[\rho(z)e^{i(kx-\omega t)}\right]$ and $\mathbf{E}(\mathbf{x},t) = \mbox{Re}\left[\mathbf{E}(z)e^{i(kx-\omega t)}\right]$ without loss of generality. Here Re/Im takes the real/imaginary part of a quantity, $k\geq 0$ denotes the wavenumber while $\omega$ is the SPW eigen-frequency to be determined by the equation of motion. In general, $\omega$ can be complex. Instead of $\rho(z)$, it proves more convenient to work directly with $$\rho_q = \int^{\infty}_0 dz~\rho(z)~\cos(qz).$$ In order for the jellium model to be valid, we must impose that $\rho_{q>q_c}=0$, where $q_c$ is a cut-off. $q^{-1}_c$ roughly gives the microscopic lattice constant of the metal. We may choose $q_c \sim k_F = mv_F/\hbar\sim n^{1/3}_0\sim k_s = \omega_{s0}/v_F$, where $k_F$ is the Fermi wavelength and $n_0$ the mean electron density. The results do not depend on the exact value of $q_c$ as long as it is sufficiently large. By definition, $$\rho(z) = \frac{2}{\pi} \int^{\infty}_0 dq ~ \rho_q~\cos(qz).$$ In terms of $\rho_q$, we can easily find, by solving Laplace's equation, that
\begin{eqnarray}
E_x(z) &=& -i\int^{\infty}_0dq~\frac{4~k~\rho_q}{k^2+q^2}\left[2\cos(qz)-e^{-kz}\right], \label{ex}\\
E_z(z) &=& \int^{\infty}_0dq~\frac{4~k~\rho_q}{k^2+q^2}\left[2(q/k)\sin(qz)-e^{-kz}\right].\label{ez} 
\end{eqnarray}
A snapshot of $\mathbf{E}(\mathbf{x},t)$ for a typical SPW is displayed in Fig.~\ref{figure:f1} (a) and the magnitude of $\mathbf{E}(z)$ is plotted in (b). Of course, in making these plots $\rho_q$ has been determined by the equation of motion to be set up shortly. 

\begin{figure*}
\begin{center}
\includegraphics[width=0.98\textwidth]{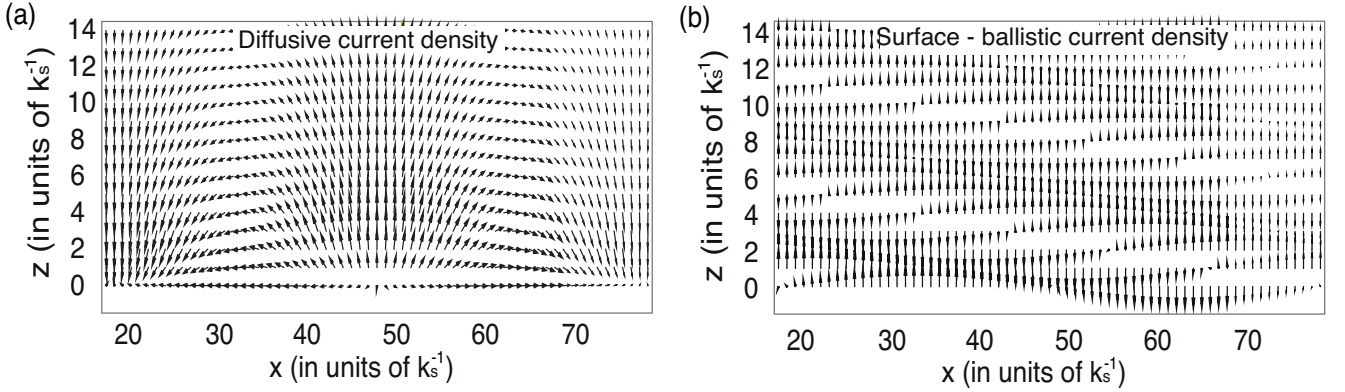}
\end{center}
\caption{Snapshots of (a) diffusive current density $\mathbf{J}_{D}(\mathbf{x},t)$ and (b) surface-ballistic current density $\mathbf{J}_B(\mathbf{x},t)$ in SPWs. These two current densities are not discriminated by the value of $\tau$, but by their dependence on the surface. $\mathbf{J}_B$ signifies genuine surface effects and would totally disappear without the surface, while $\mathbf{J}_D$ is a bulk property. $k/k_s = 0.1$ and $p=1$. Im$(\bar{\omega})$ has been neglected to emphasize the differences. \label{figure:f2}}
\end{figure*}

\textit{Current densities} - Under $\mathbf{E}(\mathbf{x},t)$, an electrical current, of density $\mathbf{J}(\mathbf{x},t) = \mbox{Re}\left[\mathbf{J}(z)e^{i(kx-\omega t)}\right]$, will flow. It can be calculated using Boltzmann's transport equation, see Sec.~\ref{sec:3}. As a crucial observation, we find that $\mathbf{J}(\mathbf{x},t)$ can always be written in two disparate contributions, $\mathbf{J}(z) = \mathbf{J}_D(z) + \mathbf{J}_B(z)$. What sets them apart is their distinct relations to $\mathbf{E}(z)$, as illustrated in Fig.~\ref{figure:f2}. It turns out that $\mathbf{J}_D(z)$ follows $\mathbf{E}(z)$ almost locally, as in the conventional hydrodynamic/Drude model -- which is applicable only when electrons execute diffusive motions, i.e. $\tau$ is very small. Explicitly, $\mathbf{J}_D(z)$ can be written 
\begin{equation}
\mathbf{J}_D(z) = \frac{i}{\bar{\omega}}\frac{\omega^2_p}{4\pi} \mathbf{E}(z) + \mathbf{J}'(z), \label{jd}
\end{equation}
where $\bar{\omega} = \omega + i/\tau$, $\omega_p = \sqrt{4\pi n_0e^2/m}$ is the characteristic plasma frequency of the metal and 
\begin{equation}
\mathbf{J}'(z) = \int^{\infty}_0 dq \frac{8 \rho_q~\mathbf{F}(k,q;\bar{\omega})}{k^2+q^2} \cos(qz),  
\end{equation}
signifies a non-local contribution that generates dispersive plasma waves, with $$\mathbf{F}(k,q;\bar{\omega}) = \left(\frac{m}{2\pi\hbar}\right)^3
 \int d^3\mathbf{v} (-e^2f'_0)~\mathbf{v}\sum^{\infty}_{l=2}\left(\frac{kv_x+qv_z}{\bar{\omega}}\right)^l.$$
Here $f_0(\varepsilon)$ denotes the Fermi-Dirac distribution and $f'_0$ its derivative, which is to be taken at zero temperature in this paper. Only terms with odd $l$ in the series contribute. One can show that $J'_z(0)\equiv 0$. The factor heading $\mathbf{E}(z)$ in Eq.~(\ref{jd}) is recognized as the Drude conductivity. As this conductivity is nearly imaginary, $\mathbf{J}_D(\mathbf{x},t)$ points almost perpendicular to $\mathbf{E}(\mathbf{x},t)$, as seen in Fig.~\ref{figure:f2} (a).  

In contrast, $J_B(z)$ -- the 'surface-ballistic' current -- has no simple relation with $\mathbf{E}(z)$ and reflects genuine surface effects: $\mathbf{J}_B(z)$ would totally disappear if the surface were absent. While $\mathbf{J}_D(z)$ is essentially a bulk property, $\mathbf{J}_B(z)$ originates from the interplay between ballistic motions and the surface. It consists of two parts, $$\mathbf{J}_B(z) = \mathbf{J}_{B,emg}(z) + p~\mathbf{J}_{B,ref}(z),$$ where $\mathbf{J}_{B,emg}$(z) is contributed by electrons directly emerging at the surface while $\mathbf{J}_{B,ref}(z)$ by reflected electrons. Here $p$ is the \textit{Fuchs} parameter, which gauges the probability that an electron impinging on the surface gets reflected back. Both contributions arise from electrons moving away from the surface, $v_z\geq 0$. In fact, 
\begin{equation}
\mathbf{J}_{B,emg/ref}(z) = \left(\frac{m}{2\pi\hbar}\right)^3\int d^3\mathbf{v}~\Theta(v_z)~e\mathbf{v}~e^{i\frac{\tilde{\omega}z}{v_z}} g_{B,emg/ref}(\mathbf{v}). \label{5}
\end{equation}
where $\tilde{\omega} = \bar{\omega} - kv_x$ and 
\begin{eqnarray}
g_{B,emg/ref}(\mathbf{v}) = -ef'_0 \int^{\infty}_{0} dq\frac{4\rho_qL_{emg/ref}(v_x,v_z,k,q;\bar{\omega})}{k^2+q^2}, \label{g}
\end{eqnarray}
with $$L_{emg/ref}(v_x,v_z,k,q;\bar{\omega}) = \frac{k(iv_x-v_z)}{kv_z\pm i\tilde{\omega}}-\frac{2(q^2v^2_z\pm \tilde{\omega}kv_x)}{\tilde{\omega}^2-q^2v^2_z}.$$
For $p\approx 1$ and small $kv_F/\bar{\omega}$, it is easy to show that $J_{B,x}(z) \approx 0$, implying that $\mathbf{J}_{B}(z)$ points normal to the surface. This is illustrated in Fig.~\ref{figure:f2} (b). 

\textit{Positiveness of $\gamma_0$} - Equation (\ref{5}) strongly constrains the value of $\bar{\omega}$. Actually, it requires that Im$(\bar{\omega})\geq 0$; otherwise, the integral in this equation would diverge for any $z$, because $e^{i\frac{\tilde{\omega}z}{v_z}}$ diverges exponentially for $v_z\rightarrow 0$. In Sec.~\ref{sec:3} and Appendix~\ref{sec:edf}, this point is further elaborated. Since 
\begin{equation}
\gamma = \mbox{Im}(\omega) = \gamma_0 - \frac{1}{\tau}, \quad \gamma_0=\mbox{Im}(\bar{\omega}),
\end{equation}
the fact that Im$(\bar{\omega})\geq 0$ suggests an intrinsic amplification channel $\gamma_0$ competing with the loss channel $\tau^{-1}$. In what follows, we show that $\bar{\omega}$ is independent of $\tau^{-1}$ by means of the equation of motion. 

\textit{Equation of motion} - The equation of motion can be obtained by relating $\rho(z)$ and $\mathbf{J}(z)$ via the equation of continuity, 
\begin{equation}
-i\bar{\omega}\rho(z) + \nabla\cdot\mathbf{J}(z) = -J_z(0)\delta(z), \label{eocc}
\end{equation}
which is derived and discussed in the next section. Here $\nabla = (ik,\partial_y,\partial_z)$. Substituting $\mathbf{J}(z)$ into this equation and Fourier transforming it, we arrive at
\begin{equation}
\int^{\infty}_0dq'\left[\mathcal{H}(q,q';\bar{\omega})-\bar{\omega}^2\delta(q-q')\right]\rho_{q'} = S(\bar{\omega}), \label{eom}
\end{equation} 
where $S(\bar{\omega})= i\bar{\omega}J_z(0)$ counts as a source term and $$\mathcal{H}(q,q';\bar{\omega}) = \Omega^2(k,q;\bar{\omega})\delta(q-q')+\mathcal{M}(q,q';\bar{\omega})$$ are the elements of a matrix denoted by $\mathcal{H}(\bar{\omega})$, with
\begin{equation}
\Omega^2(k,q;\bar{\omega}) = \omega^2_p + \frac{4\pi \bar{\omega} ~\mathbf{k}\cdot\mathbf{F}(k,q;\bar{\omega})}{\mathbf{k}\cdot\mathbf{k}}, \quad \mathbf{k} = (k,q). \label{Omega}
\end{equation}
See that $\Omega(k,q;\bar{\omega})$ is an even function of $\bar{\omega}$ and depends on the length but not the direction of $\mathbf{k}$. The matrix $\mathcal{M}(\bar{\omega})$ arises from $\nabla\cdot\mathbf{J}_{B}(z)$ with elements $\mathcal{M}(q,q';\bar{\omega})$ given in Sec.~\ref{sec:3}. In Appendix~\ref{sec:M}, we show that $\mathcal{M}(\bar{\omega}) \approx M_0\mathbb{Z}$, where $M_0 \sim kv_F/\omega_p$ is a constant and $\mathbb{Z}$ is the unity matrix will all elements being one. It makes only a minor correction to the diagonal matrix with elements $\Omega(k,q;\bar{\omega})\delta(q-q')$. 

The equation of motion (\ref{eom}) can now be re-cast in a compact matrix form,
\begin{equation}
\left[\mathcal{H}(\bar{\omega}) - \bar{\omega}^2\mathbb{I}\right] \rho = S(\bar{\omega})~\mathbb{E}, \label{eomb}
\end{equation}
where $\mathbb{I}$ is the identity matrix, $\rho$ is a column vector collecting all $\rho_q$ and $\mathbb{E}$ is a column vector with all elements being one. For later use, the source term can be rewritten 
\begin{equation}
S(\bar{\omega}) = \mathcal{G}^{t}(\bar{\omega})\rho = \int^{\infty}_0 dq~\mathcal{G}_q\rho_q, \quad \mathcal{G}_q = \frac{4~G(k,q;\bar{\omega})}{k^2+q^2}, \label{s}
\end{equation}
where $\mathcal{G}(\bar{\omega})$ is another column vector, $t$ takes the transpose, and $$G(k,q;\bar{\omega}) = G_D(k)+G_B(k,q;\bar{\omega}),$$ with $$G_D(k) = k ~ \frac{\omega^2_p}{4\pi}$$ arising from $J_{D,z}(0)$ and $$G_B(k,q;\bar{\omega}) = G_{B,emg}(k,q;\bar{\omega}) + pG_{B,ref}(k,q;\bar{\omega})$$ from $J_{B,z}(0)$. Here 
\begin{eqnarray}
&~&G_{B,emg/ref}(k,q;\bar{\omega}) = i\bar{\omega}\left(\frac{m}{2\pi\hbar}\right)^3 \nonumber \\ &~& ~\quad \times  \int d^3\mathbf{v} \Theta(v_z)(-e^2f'_0)v_z~L_{emg/ref}(v_x,v_z,k,q;\bar{\omega}). \label{gb}
\end{eqnarray}
Equations (\ref{eom}) - (\ref{gb}) are exact and do not explicitly involve $\tau$, thereby concluding that $\bar{\omega}$ is independent of $\tau$. 

\textit{Bulk modes and localized modes} - Without the surface we have $\mathcal{M}(q,q';\bar{\omega})\equiv 0$ and $S(\bar{\omega}) = 0$. Equation (\ref{eom}) reduces to $\Omega^2(k,q;\bar{\omega}) - \bar{\omega}^2 = 0$ for any $\rho_q\neq 0$, which all are extended modes and therefore describe cosine bulk plasma waves. If we discretize $q$ in step $dq$, in total there are $N_c = q_c/dq$ such modes. Solving this equation, we find the dispersion relation $\omega_b(\mathbf{k})$ for these modes. Generally, $\Omega(k,q;\bar{\omega})$ could possess an imaginary part due to a pole, located at $\bar{\omega}=kv_x+qv_z$, of the integrand in the integral involved in $\mathbf{F}(k,q;\bar{\omega})$, giving rise to the celebrated Landau damping. Neglecting this, one obtains for small $~\abs{\mathbf{k}}$
\begin{equation}
\omega^2_b(\mathbf{k}) \approx \omega^2_p ~\left[1+\frac{3}{5}\frac{\mathbf{k}\cdot\mathbf{k}v^2_F}{\omega^2_p}\right] \approx \omega^2_p, \label{hydro}
\end{equation}
which was well known from the hydrodynamic/Drude model. 

In the presence of the surface, equation (\ref{eom}) admits not only extended modes, for which $S(\bar{\omega})=0$, but also localized modes, for which $S(\bar{\omega})\neq 0$. The number of total modes cannot change and is still $N_c$. The extended modes again represent bulk waves and for them equation (\ref{eomb}) reduces to $\left[\mathcal{H}(\bar{\omega}) - \bar{\omega}^2\mathbb{I}\right]\rho = 0$. As shown in Appendix \ref{sec:M}, because of the constraint that $S(\bar{\omega})=0$, there are in total at most $N_c-1$ solutions to this equation. The number of total bulk modes is then reduced by one from that without the surface. Their dispersion relation is negligibly affected by $\mathcal{M}(\bar{\omega})$ and still given by $\omega_b(\mathbf{k})$. 

The missing bulk mode has been converted into a localized mode satisfying $S(\bar{\omega})\neq0$ and representing SPWs, for which equation (\ref{eomb}) yields
\begin{equation}
\rho = S(\bar{\omega}) \left[\mathcal{H}(\bar{\omega}) - \bar{\omega}^2\mathbb{I}\right]^{-1}\mathbb{E}. \label{rho}
\end{equation}
Plugging this in Eq.~(\ref{s}), we obtain 
\begin{equation}
1 = \mathcal{G}^{t}(\bar{\omega})\left[\mathcal{H}(\bar{\omega}) - \bar{\omega}^2\mathbb{I}\right]^{-1}\mathbb{E},
\end{equation}
which determines $\bar{\omega}$ and hence the SPW eigen-frequency $\omega$. Upon omitting $\mathcal{M}(\bar{\omega})$ from $\mathcal{H}(\bar{\omega})$, this equation becomes
\begin{equation}
1 = \int^{\infty}_0 dq~\frac{4~G(k,q;\bar{\omega})}{k^2+q^2}~\frac{1}{\Omega^2(k,q;\bar{\omega}) - \bar{\omega}^2}. \label{spw}
\end{equation}
Let us write the solution as $\bar{\omega} = \omega_s+i\gamma_0$ and hence $\omega = \omega_s+i\gamma$ with $\gamma = \gamma_0-1/\tau$. See that the solutions $\pm\omega_s+i\gamma_0$ occur together, in accord with the fact that $\rho(\mathbf{x},t)$ is a real-valued field. In Eq.~(\ref{spw}), $\Omega(k,q;\bar{\omega})$ is generally complex and $\gamma_0$ automatically includes Landau damping. 

\textit{Hydrodynamic/Drude limits} - The hydrodynamic model is revisited if we replace in Eq.~(\ref{spw}) $G(k,q;\bar{\omega})$ and $\Omega(k,q;\bar{\omega})$ with $G_D(k)$ and $\omega_b(\mathbf{k})$ given by Eq.~(\ref{hydro}), respectively. If we further disregard the dispersion in $\omega_b(\mathbf{k})$, the Drude model is then recovered, in which case we immediately find $\omega_s = \omega_{s0}$. In both models, $\bar{\omega}$ is real. 

\textit{Approximate solutions} - We can solve Eq.~(\ref{spw}) approximately. To the lowest order in $\gamma_0/\omega_s$, which is assumed to be small, we may determine $\omega_s$ by approximating the real part of (\ref{spw}) as follows
\begin{equation}
1 \approx \int^{\infty}_0 dq~\frac{4~\mbox{Re}\left[G(k,q;\omega_s)\right]}{k^2+q^2}~\frac{1}{\Omega^2(k,q;\omega_s) - \omega^2_s}. \label{18}
\end{equation}
Substituting the so-obtained $\omega_s$ in the imaginary part of (\ref{spw}), we find
\begin{equation}
\frac{\gamma_0}{\omega_s} \approx -\frac{1}{2}\frac{\int^{\infty}_0 dq~\frac{4~\mbox{Im}\left[G(k,q;\omega_s)\right]}{k^2+q^2}~\frac{1}{\Omega^2(k,q;\omega_s) - \omega^2_s}}{ \int^{\infty}_0 dq~\frac{4~\mbox{Re}\left[G(k,q;\omega_s)\right]}{k^2+q^2}~\frac{1}{\Omega^2(k,q;\omega_s) - \omega^2_s} \frac{\omega^2_s}{\Omega^2(k,q;\omega_s)-\omega^2_s}},
\end{equation}
which can be brought into a rather simple form if we take $\Omega(k,q;\omega_s)\approx \omega^2_p$ and $\omega_s\approx \omega_{s0}$. Namely, 
\begin{equation}
\frac{\gamma_0}{\omega_s} \approx -\frac{1}{2}\frac{\int^{\infty}_0 dq~\frac{\mbox{Im}\left[G(k,q;\omega_s)\right]}{k^2+q^2}}{\int^{\infty}_0 dq~\frac{\mbox{Re}\left[G(k,q;\omega_s)\right]}{k^2+q^2}} = \frac{1}{2}\frac{\mbox{Re}\left[J_z(0)\right]}{\mbox{Im}\left[J_z(0)\right]}. \label{gamma0}
\end{equation} 
In this approximation, $\Omega(k,q;\omega_s)$ could have an imaginary part only if $q_c>\omega_s/v_F$. Landau damping would be excluded from $\gamma_0$ otherwise. 

To proceed, we need to evaluate $G_B(k,q;\bar{\omega})$. To the linear order in $k$, we find
\begin{eqnarray}
&~&G_{B}(k,q;\bar{\omega}) \approx i\bar{\omega}e^2\left(\frac{m}{2\pi\hbar}\right)^3 \nonumber \\ &~& ~ \times  \int d^3\mathbf{v} \Theta(v_z)f'_0v_z\left[\frac{2q^2v^2_z(1+p)}{\bar{\omega}^2-q^2v^2_z} + \frac{kv_z(1-p)}{i\bar{\omega}}\right]. \label{gbb}
\end{eqnarray}
Carrying out the integral gives 
\begin{equation}
G_B(k,q;\bar{\omega}) \approx \frac{\omega^2_p}{4\pi}\left[\frac{k(p-1)}{2} - i \frac{3(1+p)}{4}\frac{v_F}{\bar{\omega}}q^2\right]. 
\end{equation}
Thus, Re$\left[G(k,q;\omega_s)\right] \approx k(\omega^2_p/4\pi)(1+p)/2$. Inserting this in Eq.~(\ref{18}), we get $\omega_s \approx \omega_{s0}\sqrt{2-\frac{p+1}{2}}. $
We see that $\omega_s$ generally depends on surface scattering effects, in contrast with what is expected of the hydrodynamic/Drude model. Analogously, using Eq.~(\ref{gamma0}) and $q_cv_F \sim \omega_{s0}$, we find $\gamma_0 \approx \alpha \omega_{s0}$, with $\alpha = 3/2\pi$. Landau damping has been excluded here, as the approximation only takes the real part of $\Omega(k,q;\omega_s)$. 

\textit{Numerical solutions} - Equation (\ref{spw}) can also be exactly solved numerically. A comparison with the approximate solution is displayed in Fig.~\ref{figure:f0}. The agreement in the matter of $\omega_s$ is satisfactory, while that for $\gamma_0$ is not. The discrepancy might be because the approximate solution excludes Landau damping. It should be emphasized that, our numerical solutions do not depend on the choice of $q_c$, provided the latter is big enough, i.e. $q_c\geq k_s$. 
 
\section{Methods}
\label{sec:3}
This section discusses further some technical aspects of the theory. We set out with a discussion of the equation of continuity in the presence of surfaces. Then we describe Boltzmann's equation and solve it to obtain the electronic distribution functions. Thence we derive the dynamic equation for the charge density. 

\textit{Equation of continuity} - We start with the equation of continuity, $\left(\partial_t+1/\tau\right)\rho(\mathbf{x},t)+\partial_{\mathbf{x}}\cdot\mathbf{j}(\mathbf{x},t)=0$, which relates the charge density $\rho(\mathbf{x},t)$ and the current density $\mathbf{j}(\mathbf{x},t)$ in a generic way. Here the damping term $-\rho(\mathbf{x},t)/\tau$ is included to account for electronic collisions that would drive the system toward equilibrium. In the jellium model, $\rho(\mathbf{x},t)$ appears when the electron density is disturbed away from its equilibrium value $n_0$. The surface  prevents any electrons from leaking out of the metal. Explicitly, we write $\mathbf{j}(\mathbf{x},t) = \Theta(z)\mathbf{J}(\mathbf{x},t)$, where $\Theta(z)$ is the Heaviside step function. In so doing, we have treated the surface as a hard wall and considered the fact that $\mathbf{J}(\mathbf{x},t)$ may not vanish even in the immediate neighborhood of the surface -- which is obviously the case with the hydrodynamic/Drude model. The equation of continuity becomes
\begin{equation}
\left(\partial_t+\frac{1}{\tau}\right)\rho(\mathbf{x},t)+\partial_{\mathbf{x}}\cdot\mathbf{J}(\mathbf{x},t) = -J_z(\mathbf{x}_0,t)\delta(z), \label{eoc}
\end{equation} 
where $\mathbf{x}_0 = (\mathbf{r},0)$ denotes a point on the surface and $\delta(z)$ is the Dirac function peaked on the surface. Physically, the right hand side of Eq.~(\ref{eoc}) means that, charges must pile up on the surface if they do not come to a halt before they reach it. 

Equation (\ref{eoc}) reduces to equation (\ref{eocc}) upon using the plane wave form for $\rho(\mathbf{x},t)$ and $\mathbf{J}(\mathbf{x},t)$.     

\textit{Boltzmann's approach} - We employ Boltzmann's equation to calculate $\mathbf{J}(\mathbf{x},t)$ as a response to $\mathbf{E}(\mathbf{x},t)$. On the microscopic level, we may introduce a surface potential $\phi_s(\mathbf{x})$ into the equation to account for surface scattering effects. The corresponding surface field $\mathbf{E}_s(\mathbf{x}) = -\partial_{\mathbf{x}}\phi_s(\mathbf{x})$ is peaked on the surface and, complying with the hard-wall picture, may have an infinitesimal spread $d_s\rightarrow 0_+$. Unfortunately, as $\phi_s(\mathbf{x})$ can hardly be known and varies from one sample to another, this microscopic method is impractical and futile. 

Alternatively, surface scattering effects can be dealt with using boundary conditions~\cite{ziman1960,abrikosov,reuter1948,kaganov1997}. This is possible because $\mathbf{E}_s(\mathbf{x})$ acts only on the surface. In the bulk, the solutions -- the electronic distribution function $f(\mathbf{x},\mathbf{v},t)$ -- to Boltzmann's equation can be uniquely determined up to some parameters, which summarize the effects of -- while without actually knowing -- $\phi_s(\mathbf{x})$. With translational symmetry along the surface, only one such parameter, namely, the so-called \textit{Fuchs} parameter $p$, is needed for the simple specular scattering picture. Physically, $p$ measures the probability that an electron impinging upon the surface is bounced back. As usual, we write $f(\mathbf{x},\mathbf{v},t) = f_0(\varepsilon(\mathbf{v}))+g(\mathbf{x},\mathbf{v},t)$, where $g(\mathbf{x},\mathbf{v},t)$ is the non-equilibrium part of the distribution function in the presence of $\mathbf{E}(\mathbf{x},t)$. The current density can then be calculated by $\mathbf{J}(\mathbf{x},t) = (m/2\pi\hbar)^2\int d^3\mathbf{v}~e\mathbf{v}~g(\mathbf{x},\mathbf{v},t)$, where $e$ denotes the electron charge. It is worth pointing out that, as $g(\mathbf{x},\mathbf{v},t)$ is a distribution only for the bulk, the charge density is not given by $\tilde{\rho}(\mathbf{x},t) = (m/2\pi\hbar)^2\int d^3\mathbf{v}~e~g(\mathbf{x},\mathbf{v},t)$, i.e. $\rho(\mathbf{x},t)\neq\tilde{\rho}(\mathbf{x},t)$. Actually, It is easy to see that $\tilde{\rho}(\mathbf{x},t)$ satisfies $(\partial_t+1/\tau)\tilde{\rho}(\mathbf{x},t)+\partial_{\mathbf{x}}\cdot\mathbf{J}(\mathbf{x},t)=0$ rather than Eq.~(\ref{eoc}). Obviously, what is missing from $\tilde{\rho}(\mathbf{x},t)$ is exactly the charges just localized on the surface. 

\textit{Electronic distribution function} - Let us write $g(\mathbf{x},\mathbf{v},t) = $ Re $\left[g(\mathbf{v},z)e^{i(kx-\omega t)}\right]$. In the regime of linear responses, Boltzmann's equation reads
\begin{equation}
\frac{\partial g(\mathbf{v},z)}{\partial z} + \lambda^{-1}~g(\mathbf{v},z) +
 ef'_0(\varepsilon)~\frac{\mathbf{v}\cdot\mathbf{E}(z)}{v_z} = 0,  
\label{be}
\end{equation} 
where $ \lambda = iv_z/\tilde{\omega}$. In this equation, the velocity $\mathbf{v}$ is more of a parameter than an argument and can be used to tag electron beams. It is straightforward to solve the equation under appropriate boundary conditions (Appendix~\ref{sec:edf}). Naturally, we have $$g(\mathbf{v},z) = g_{bulk}(\mathbf{v},z)+g_{surface}(\mathbf{v},z),$$ where the bulk term would exist even in the absence of surfaces whereas the surface term would not. Using the expressions for $\mathbf{E}(z)$, we obtain
\begin{equation}
g_{bulk}(\mathbf{v},z) = -ef'_0\int^{\infty}_{-\infty} dq~\frac{4\rho_q}{k^2+q^2}\frac{kv_x+qv_z}{\bar{\omega}-(kv_x+qv_z)}~e^{iqz}, \label{gbl}
\end{equation}
which has the same form as one would expect for a bulk system. Here we have defined $\rho_{q<0} := \rho_{-q}$. 

As for $g_{surface}(\mathbf{v},z)$, we find it with a subtle structure: it can be written as a sum of two contributions, one of which, $g_{D,surface}(\mathbf{v},z)$, has a single form for all electrons irrespective of their velocities while the other, $g_{B,surface}(\mathbf{v},z)$, does not. Explicitly, we have
\begin{equation}
g_{D,surface}(\mathbf{v},z) = -ef'_0\int^{\infty}_{0} dq~\frac{4\rho_q}{k^2+q^2}\frac{k(v_z-iv_x)}{kv_z+i\tilde{\omega}}e^{-kz}. \label{gd}
\end{equation}
We may combine $g_{bulk}(\mathbf{v})$ and $g_{D,surface}(\mathbf{v},z)$ in a single term, $$g_D(\mathbf{v},z) = g_{bulk}(\mathbf{v},z) + g_{D,surface}(\mathbf{v},z),$$ in order to separate them from $$g_B(\mathbf{v},z):= g_{B,surface}(\mathbf{v},z).$$ In so doing, we have decomposed $$g(\mathbf{v},z) = g_D(\mathbf{v},z)+g_B(\mathbf{v},z)$$ in a diffusive and a ballistic component. It is underlined that $g_B(\mathbf{v},z)$ arises only when the surfaces are present. For bulk systems without surfaces, it does not exist even if the electronic motions are totally ballistic, i.e. $\tau\rightarrow\infty$. As such, we call it 'surface-ballistic'. 

$g_{B}(\mathbf{v},z)$ exists only for electrons leaving the surface, i.e. $v_z\geq0$. Those electrons could directly emerge from the surface or be those incident on but subsequently get reflected by the surface. What fundamentally sets $g_{B,surface}$ apart from $g_D(\mathbf{v},z)$ rest with its simple $z$ dependence. Actually, we have $$g_{B}(\mathbf{v},z) = \Theta(v_z)~e^{i\frac{\tilde{\omega}z}{v_z}}\left[g_{B,emg}(\mathbf{v})+p~g_{B,ref}(\mathbf{v})\right],$$ where the \textit{Fuchs} parameter $p$ gauges the fraction of reflected electrons and $g_{B,emg/ref}(\mathbf{v})$ has been given in Eq.~(\ref{g}). As already remarked, since $g_B(\mathbf{v},z) \propto e^{i\tilde{\omega}z/v_z}\propto e^{-\gamma_0z/v_z}$, we must have $\gamma_0\geq 0$; otherwise, it would diverge either when $z\rightarrow\infty$ or for small $v_z$. In Appendix~\ref{sec:edf}, we argue that this result is a consequence of the causality principle, which states that the number of reflected electrons is determined by that of incident electrons, rather than otherwise.  

\textit{Divergence of the current densities} - The current density $\mathbf{J}(z) = (m/2\pi\hbar)^2\int d^3\mathbf{v} ~e\mathbf{v}~g(\mathbf{v},z)$ is split in two parts: $\mathbf{J}_D(z)$ and $\mathbf{J}_B(z)$, where $\mathbf{J}_{D/B}(z) = (m/2\pi\hbar)^2\int d^3\mathbf{v} ~e\mathbf{v}~g_{D/B}(\mathbf{v},z)$. Using the expressions for $g_{D/B}(\mathbf{v},z)$, it is straightforward to obtain $\mathbf{J}_D(z)$ given by Eq.~(\ref{jd}) and $\mathbf{J}_B(z)$ by Eq.~(\ref{5}). To obtain the equation of motion (\ref{eom}) from the equation of continuity (\ref{eocc}), the Fourier transform of $\nabla\cdot\mathbf{J}_{D/B}(z)$ is needed. We find
\begin{equation}
\int^{\infty}_0dz~\cos(qz)~\nabla\cdot \mathbf{J}_D(z) = \frac{i}{\bar{\omega}} \Omega^2(k,q;\bar{\omega}) \rho_q, \label{dd}
\end{equation}
with $\Omega(k,q;\bar{\omega})$ given by Eq.~(\ref{Omega}). Additionally, 
\begin{equation}
\int^{\infty}_0 dz \cos(qz) \mathbf{\nabla}\cdot \mathbf{J}_B(z) = \frac{i}{\bar{\omega}} \int^{\infty}_0dq'~\mathcal{M}(q,q')~\rho_{q'}, \label{30}
\end{equation}
where $\mathcal{M}(q,q')$ is a matrix given by
\begin{eqnarray}
&~&\mathcal{M}(q,q') = \frac{4\bar{\omega}^2}{k^2+q^{'2}}\left(\frac{m}{2\pi\hbar}\right)^3\nonumber\\ &~& ~ \times\int d^3\mathbf{v}\Theta(v_z)\left(-e^2f'_0\right)\frac{i\tilde{\omega}v_z}{\tilde{\omega}^2-q^2v^2_z}~L(v_x,v_z,k,q';\bar{\omega}),
\end{eqnarray}
where $$L(v_x,v_z,k,q;\bar{\omega}) = L_{emg}(v_x,v_z,k,q;\bar{\omega}) + pL_{ref}(v_x,v_z,k,q;\bar{\omega}).
$$ In Appendix~\ref{sec:M}, we show that $\mathcal{M}(q,q')$ generally makes a minor correction, of the order of $kv_F/\omega_s$, to $\Omega^2(k,q;\bar{\omega})\delta(q-q')$. Ultimately, the insignificance of this correction may be ascribed to the factor $e^{i\tilde{\omega}z/v_z}$ in $g_B(\mathbf{v},z)$, which is extremely oscillatory over $z$ with a period less than $\sim v_F/\omega_s$.

Substituting Eqs.~\ref{dd} and (\ref{30}) in the equation of continuity, we immediately obtain the equation of motion (\ref{eom}). 

\section{discussions and conclusions}
\label{sec:4}
Thus, on the basis of Boltzmann's equation, we have established a rigorous theory for SPWs in metals with arbitrary electronic collision rate $1/\tau$. As a key consequence of the theory, we find that there exists a self-amplification channel for SPWs, which would cause the latter to spontaneously amplify at a rate $\gamma_0$ if not for electronic collisions. Surprisingly, the value of $\gamma_0$ turns out to be independent of $\tau$. The presence of this channel is guaranteed by the causality principle.  

Whether the system could actually amplify or not depends on the competition between $\gamma_0$ and $1/\tau$. If $\gamma_0>1/\tau$, SPWs will amplify and the system will become unstable. In our theory, the non-equilibrium deviation $g(\mathbf{v},z)$ refers to the Fermi-Dirac distribution $f_0(\varepsilon)$; as such, the instability is one of the Fermi sea. Needless to say, the instability will be terminated once the system deviates far enough from the Fermi sea and settles in a stable state. Clarifying the nature of the destination state is a subject of crucial importance for future study. 

One central feature of our theory is the classification of current densities into a diffusive component $\mathbf{J}_D(z)$ and a surface-ballistic component $\mathbf{J}_B(z)$. This classification is not based on the value of $\tau$ but according to whether the component obeys the (generalized) Ohm's law or not. Apart from this, these components are also discriminated in other ways. Firstly, they are controlled by different length scales. As it largely follows the local electric field $\mathbf{E}(z)$, the characteristic length associated with $\mathbf{J}_D(z)$ is $k^{-1}$. On the other hand, the length for $\mathbf{J}_B(z)$ is $v_F/\gamma_0$, because of its simple $z$-dependence. Secondly, they are oriented disparately. $\mathbf{J}_D(z)$ is largely oriented normal to $\mathbf{E}(z)$ locally whereas $\mathbf{J}_B(z)$ normal to the surfaces -- especially for $p$ close to unity. Thirdly, $\mathbf{J}_D(z)$ is a bulk property and exists regardless of the surface; On the contrary, $\mathbf{J}_B(z)$ reflects true surface effects and it would disappear without surfaces. 

If we replace the vacuum by a dielectric $\epsilon$, $\gamma_0$ may
be reduced by an order of ~$1/\epsilon$ due to weakened $E_z(z)$ and $J_z(0)$. Roughly speaking, $\rho(z)$ present on the metal side
induces mirror charges amounting to
$\rho'(z)=-\rho(-z)(\epsilon-1)/(\epsilon+1)$ on the dielectric
side. If $\rho(z)$ is highly localized about the interface, as is
with SPWs, $\rho(z)$
and $\rho'(z)$ will combine to give a net charge of
$2\rho(z)/(\epsilon+1)$. As a result, $E_z$ and $J_z(0)$ will be
reduced by a factor of $2/(\epsilon+1)$. This leads to smaller
$\gamma_0$ and smaller $\omega_{s0}=\omega_p/\sqrt{\epsilon+1}$. Studying dielectric effects may be important in applications. 

Another problem that needs to be addressed for experimental studies is concerned with the effects of inter-band transitions. In the most experimented materials, such as silver and gold, these transitions are known to have dramatic effects. They not only open a loss channel due to inter-band absorption, but also significantly shift the SPW frequency. Including them in our formalism consists of a simple generalization: in addition to $\mathbf{J}_D(z)$ and $\mathbf{J}_B(z)$, the total current density $\mathbf{J}(z)$ must now also have a component $\mathbf{J}_{int}(z)$ accounting for inter-band transitions. The equation of motion is obtained by substituting $\mathbf{J}(z)$ in Eq.~(\ref{eocc}). One may write $J_{int,\mu}(z) = \sum_{\nu}\int dz'~\sigma_{\mu\nu}(z,z';\omega)E_{\nu}(z')$, where $\mu, \nu = x,y,z$ and the inter-band conductivity $\sigma_{\mu\nu}$ can in principle be calculated using Greenwood-Kubo formula. In practice, calculating $\sigma_{\mu\nu}$ could be a formidable task even for the imaginably simplest surfaces. Nevertheless, one may argue that $\mathbf{J}_{int}(z)$ primarily affects the properties of bulk waves, namely, $\Omega(k,q;\bar{\omega})$. The causality principle should still protect the amplification channel. A systematic analysis will be presented elsewhere. 

We remark that, $\gamma_0$ can also be calculated by studying the temporal evolution of the electrostatic potential energy of the system. In particular, equation (\ref{gamma0}) can be directly derived in this way. Detailed calculations along this line will be published in a separate paper. 
 
To conclude, we have presented a theory for SPWs taking into account the unique interplay between ballistic electronic motions and boundary effects, from which it emerges a universal self-amplification channel for these waves. It is expected that the study will bear far-reaching practical and fundamental consequences, to be explored in the future. 
\\
\\
\noindent
\textbf{Acknowledgement} -- HYD acknowledges the International Research
Fellowship of the Japan Society for the Promotion of Science
(JSPS). This work is supported by JSPS KAKENHI Grant Nos. 15K13507 and 15K21722 and
MEXT KAKENHI Grant No. 25107005.

\appendix
\section{Electronic distribution functions}
\label{sec:edf}
The general solution to Eq.~(\ref{be}) is given by 
\begin{equation}
g(\mathbf{v},z) = e^{i\frac{\tilde{\omega}z}{v_z}}\left(C(\mathbf{v})-\frac{e\partial_{\mathbf{v}}f_0}{mv_z}\cdot \int^z_0~dz'~e^{-i\frac{\tilde{\omega}z'}{v_z}}\mathbf{E}(z')\right), \label{b1}
\end{equation}
where $C(\mathbf{v})$ is an arbitrary integration constant to be determined by boundary conditions. We require $g(\mathbf{v},z)=0$ at $z\rightarrow\infty$ for any $\mathbf{v}$. For electrons moving away from the surface, i.e. $v_z>0$, this condition is fulfilled for any $C(\mathbf{v})$. For electrons moving toward the surface, i.e. $v_z<0$, we may choose 
\begin{equation}
C(\mathbf{v}) = \frac{e\partial_{\mathbf{v}}f_0}{mv_z}\cdot \int^{\infty}_0~dz'~e^{-i\frac{\tilde{\omega}z'}{v_z}}\mathbf{E}(z'), \quad v_z<0,
\end{equation}
which leads to 
\begin{equation}
g(\mathbf{v},z) = \frac{e\partial_{\mathbf{v}}f_0}{mv_z}\cdot \int^{\infty}_zdz' ~e^{i\frac{\tilde{\omega}(z-z')}{v_z}}\mathbf{E}(z'), \quad v_z<0. 
\end{equation}
To determine $C(\mathbf{v})$ for $v_z>0$, the boundary condition at $z=0$ has to be used, which depends on surface properties. In this paper, we assume that a fraction $p$ (\textit{Fuchs} parameter) of the electrons impinging on the surface are bounced back in the absence of $E_z(z)$, i.e. $$g\left((v_x,v_y,v_z>0),z=0\right)=p~g\left((v_x,v_y,-v_z),z=0\right)$$ evaluated at $E_z(z)=0$. Then we get 
\begin{equation}
C(\mathbf{v}) = - p~\frac{e\partial_{\mathbf{v}}f_0}{mv_z}\cdot \int^{\infty}_0dz' e^{i\frac{\tilde{\omega}(z+z')}{v_z}}\mathbf{E}(z'), \quad v_z>0.
\end{equation}
By definition, $p$ varies from zero to unity. Thus, we obtain $$g(\mathbf{v},z) = \Theta(v_z)g_{>}(\mathbf{v},z) +
 \Theta(-v_z)g_{<}(\mathbf{v},z),$$ where 
 \begin{widetext}
 \begin{eqnarray}
g_{>}(\mathbf{v},z) = - \left[\int^{z}_0 e^{i\tilde{\omega}|\frac{z-z'}{v_z}|}
	     + p\int^{\infty}_0
	     e^{i\tilde{\omega}|\frac{z+z'}{v_z}|}\right]
\frac{e\mathbf{v}\cdot\mathbf{E}(z')}{|v_z|}\frac{\partial f_0}{\partial
 \varepsilon}~dz', \quad
g_{<}(\mathbf{v},z) = - \int^{\infty}_z e^{i\tilde{\omega}|\frac{z-z'}{v_z}|} \frac{e\mathbf{v}\cdot\mathbf{E}(z')}{|v_z|}\frac{\partial f_0}{\partial
 \varepsilon}~dz'.
\end{eqnarray}
Utilizing equations (\ref{ex}) and (\ref{ez}) for $\mathbf{E}(z)$ and carrying out the integral over $z'$, we find
\begin{equation}
g_{>/<}(\mathbf{v},z) = -\frac{\partial f_0\left(\varepsilon(\mathbf{v})\right)}{\partial \varepsilon}~\frac{e}{\abs{v_z}} \int^{\infty}_0 dq~\frac{4 k ~ \rho_q}{k^2+q^2} ~
g_{>/<}(\mathbf{v},z;k,q),
\end{equation}
where
\begin{eqnarray}
g_>(\mathbf{v},z;k,q) &=& \frac{v_z-iv_x}{k+i\tilde{\omega}/v_z}~e^{-kz} + \frac{2\left(qv_z\frac{q}{k}+\tilde{\omega}\frac{v_x}{v_z}\right)}{(\tilde{\omega}/v_z)^2-q^2}~\cos(qz) + i~\frac{2\left(qv_x+\tilde{\omega}\frac{q}{k}\right)}{(\tilde{\omega}/v_z)^2-q^2}~\sin(qz) \nonumber\\ 
&~& ~~~~ \quad + \left\{\left(\frac{iv_x-v_z}{k+i\tilde{\omega}/v_z} - \frac{2\left(\tilde{\omega}\frac{v_x}{v_z}+qv_z\frac{q}{k}\right)}{(\tilde{\omega}/v_z)^2-q^2}\right)+p\left(\frac{iv_x-v_z}{k-i\tilde{\omega}/v_z}+\frac{2\left(\tilde{\omega}\frac{v_x}{v_z}-qv_z\frac{q}{k}\right)}{(\tilde{\omega}/v_z)^2-q^2}\right)\right\}~e^{i\frac{\tilde{\omega}z}{v_z}},
\end{eqnarray}
and 
\begin{equation}
g_<(\mathbf{v},z;k,q) = \frac{iv_x-v_z}{k-i\tilde{\omega}/\abs{v_z}}~e^{-kz}+ \frac{2\left(q\abs{v_z}\frac{q}{k}+\tilde{\omega}\frac{v_x}{\abs{v_z}}\right)}{(\tilde{\omega}/v_z)^2-q^2}~\cos(qz) - i~\frac{2\left(qv_x+\tilde{\omega}\frac{q}{k}\right)}{(\tilde{\omega}/v_z)^2-q^2} ~ \sin(qz). \label{11}
\end{equation}
Now we can combine the terms with $\cos(qz)$ and those with $\sin(qz)$ into a single term called $g_{bulk}(\mathbf{v},z)$, while the rest into $g_{surface}(\mathbf{v},z)$. Their expressions have been given in Sec.~\ref{sec:3}. 

If the surface is not uniform, we should have a function $p(\mathbf{r})$ instead of a constant $p$. The boundary condition should then be written $g(\mathbf{x}_0,(v_x,v_y,v_z>0),t) = p(\mathbf{r}) g(\mathbf{x}_0,(v_x,v_y,-v_z),t)$. In such case, the translational symmetry along the surface is generally lost and one cannot work with a single $k$-component any more, i.e. there is scattering effects and different $k$-components are mixed. We do not consider this in this paper.  

\textit{Causality principle} - In applying the boundary conditions to obtain $C(\mathbf{v})$, we have implicitly assumed Im$(\tilde{\omega})\geq 0$; otherwise, we would find unphysical solutions that violate the principle of causality, which states that the number of out-going electrons is determined by the number of in-coming electrons, not otherwise. It is easy to show that, had we assumed Im$(\tilde{\omega})<0$, we would have found the opposite: the number of reflected electrons would be fixed while the number of incident electrons would go to infinity as $p\rightarrow 0$, which is unphysical. 

\section{The matrix $\mathcal{M}(\bar{\omega})$}
\label{sec:M}
In the first place, we show that $\mathcal{M}(\bar{\omega})/\omega^2_p \propto ikv_F/\bar{\omega} + ...$, where the ellipsis stands for higher order terms in $kv_F/\bar{\omega}$. For this purpose, let us discretize $q$ in step $dq = 2\pi/d$, i.e. $q_l = l (2\pi/d)$, where $kd\gg 1$ and $l=0,1,2,...$ takes integer values. Equation (\ref{eom}) can then be brought into the following form
\begin{equation}
\left(\Omega^2(k,q_l;\bar{\omega}) - \bar{\omega}^2\right)\rho_l + \sum_{l'}\mathcal{M}_{l,l'}(\bar{\omega}) \rho_{l'} = S(\bar{\omega}), \label{b1}
\end{equation}
where $\mathcal{M}_{l,l'}(\bar{\omega}) = (2\pi/d)\mathcal{M}(q_l,q_{l'};\bar{\omega})$ and $\rho_l = \rho_{q_l}$. Writing $\int d^3\mathbf{v} ~\Theta(v_z) = \int^{2\pi}_0d\varphi \int^{\pi/2}_0d\theta \sin\theta \int^{\infty}_0 dv^2 (v/2)$ and integrating over $v$, we find 
\begin{eqnarray}
\frac{\mathcal{M}_{l,l'}(\bar{\omega})}{\omega^2_p} = i \frac{3}{2\pi kd}\int^{2\pi}_0d\varphi \int^{\pi/2}_0d\theta \sin\theta \frac{\bar{\omega}^2\cos\theta-\bar{\omega}kv_F\sin\theta\cos\theta\cos\varphi}{(\bar{\omega}-kv_F\sin\theta\cos\varphi)^2-q^2_{l}v^2_F\cos^2\theta}\frac{k\bar{\omega}/v_F}{k^2+q^2_{l'}} L(v_F\sin\theta\cos\varphi,v_F\cos\theta,k,q_{l'},\bar{\omega}). \label{c1}
\end{eqnarray}
To the lowest order in $kv_F/\bar{\omega}$, we only need to retain $L^{(0)}$ in the expansion $L^{sym} = \sum^{\infty}_{m=0}L^{(m)}\left(kv_F/\bar{\omega}\right)^m$. Thus, 
\begin{equation}
L(v_F\sin\theta\cos\varphi,v_F\cos\theta,k,q,\bar{\omega}) \approx \frac{2q^2v^2_F\cos^2\theta}{\bar{\omega}^2-q^2v^2_F\cos^2\theta}~\left(1+p\right).
\end{equation}
Substituting this back in (\ref{c1}) and approximating
\begin{equation}
\frac{\bar{\omega}^2\cos\theta-\bar{\omega}kv_F\sin\theta\cos\theta\cos\varphi}{(\bar{\omega}-kv_F\sin\theta\cos\varphi)^2-q^2v^2_F\cos^2\theta} \approx \frac{\bar{\omega}^2\cos\theta}{\bar{\omega}^2-q^2v^2_F\cos^2\theta}, \quad \frac{q^2}{q^2+k^2} \approx 1,
\end{equation}
we arrive at 
\begin{equation}
\frac{\mathcal{M}_{l,l'}(\bar{\omega})}{\omega^2_p} = - i ~ \frac{3(1+p)}{\pi kd}\left(\frac{kv_F}{\bar{\omega}}\right)\int^{2\pi}_0d\varphi \int^{\pi/2}_0d\theta \sin\theta~ \frac{\cos^3\theta}{1-(q_{l}v_F/\bar{\omega})^2\cos^2\theta}  \frac{1}{1-(q_{l'}v_F/\bar{\omega})^2\cos^2\theta}. 
\end{equation}
Obviously, we have $\mathcal{M}(\bar{\omega})/\omega^2_p \sim kv_F/\bar{\omega}$, as stated.

We may proceed further If we take
\begin{equation}
\frac{3}{\pi} \int^{2\pi}_0d\varphi \int^{\pi/2}_0d\theta \sin\theta \frac{\cos^3\theta}{1-(q_{l}v_F/\bar{\omega})^2\cos^2\theta}  \frac{1}{1-(q_{l'}v_F/\bar{\omega})^2\cos^2\theta} \approx \frac{3}{\pi} \int^{2\pi}_0d\varphi \int^{\pi/2}_0d\theta \sin\theta ~\cos^3\theta \sim 1,
\end{equation}
from which it follows that $\mathcal{M}(\bar{\omega})_{l,l'} \approx M_0 = -i\omega^2_p(1/kd)(kv_F/\bar{\omega})$, which is a constant. We then write $\mathcal{M}(\bar{\omega}) \approx M_0\mathbb{Z}$, where $\mathbb{Z}_{l,l'}=1$ constitutes a unity matrix. To simplify the following discussions, let us take $\Omega(k,q;\bar{\omega}) = \omega_p$. Equation (\ref{b1}) is now written 
\begin{equation}
\left[(\omega^2_p - \bar{\omega}^2)\mathbb{I}+M_0\mathbb{Z}\right]\rho = S(\bar{\omega}) \mathbb{E}. 
\end{equation}
For bulk modes, $S(\bar{\omega}) = 0$. Note that $\mathbb{Z}$ has only one non-vanishing eigenvalue amounting to its dimension $N_c = q_cd/2\pi \sim (\omega_s/v_F)(kd/2\pi)$. The corresponding eigenvector is $\rho \propto \mathbb{E}$, which obviously does not satisfy $S(\bar{\omega}) = 0$ and is not a bulk mode. We therefore conclude that there are in total at most $N_c-1$ bulk modes. For these modes, the eigenvalues of $\mathbb{Z}$ are all zero and therefore $\mathcal{M}(\bar{\omega})$ has no impact on bulk modes. 

We can also show that $\mathcal{M}(\bar{\omega})$ is negligible for the localized mode, for which $S(\bar{\omega})\neq0$. To this end, we write
\begin{equation}
\left[\left(\omega^2_p-\bar{\omega}^2\right)\mathbb{I}+M_0\mathbb{Z}\right]^{-1} = U^{-1}\left[\left(\omega^2_p-\bar{\omega}^2\right)\mathbb{I}+M_0\tilde{\mathbb{Z}}\right]^{-1}U,
\end{equation}
where $U$ is a similarity transformation that diagnolizes $\mathbb{Z}$. We have used a tilde to indicate the transformed matrices, e.g. we write $\tilde{\mathbb{Z}} = U\mathbb{Z}U^{-1}$. As said, $\tilde{\mathbb{Z}}$ has only one non-vanishing element. Let it be the $l_0$-th element. Then $\tilde{\mathbb{Z}}_{l,l'} = N_c\delta_{l,l_0}\delta_{l',l_0}$. See that $M_0\sim 1/N$ and $\tilde{\mathcal{M}}_{l,l'}(\bar{\omega})\sim -i(\omega^2_p/2\pi) \delta_{l,l_0}\delta_{l',l_0}$. Introducing $\tilde{\mathcal{G}}^{t}=\mathcal{G}^{t}U^{-1}$ and $\tilde{\mathbb{E}} = U\mathbb{E}$, we can rewrite the secular equation for the localized mode as 
\begin{eqnarray}
1 = \tilde{\mathcal{G}}^{t} \left[\left(\omega^2_p-\bar{\omega}^2\right)\mathbb{I}+M_0\tilde{\mathbb{Z}}\right]^{-1} \tilde{\mathbb{E}}.
\end{eqnarray}
Explicitly,
\begin{eqnarray}
1 = \sum_l \tilde{\mathcal{G}}^{t}_l \frac{1}{\omega^2_p-\bar{\omega}^2+M_0\tilde{\mathbb{Z}}_{l,l}}\tilde{\mathbb{E}}_{l} =  \sum_l \mathcal{G}^{t}_l \frac{1}{\omega^2_p-\bar{\omega}^2}\mathbb{E}_{l} + \left[\tilde{\mathcal{G}}^{t}_{l_0} \frac{1}{\omega^2_p(1+i/2\pi)-\bar{\omega}^2}\tilde{\mathbb{E}}_{l_0} - \mathcal{G}^{t}_{l_0}\frac{1}{\omega^2_p-\bar{\omega}^2}\mathbb{E}_{l_0}\right] \approx \sum_l \mathcal{G}^{t}_l \frac{1}{\omega^2_p-\bar{\omega}^2}\mathbb{E}_{l}. \nonumber
\end{eqnarray}
The term in the square bracket makes only a contribution of the order of $\sim 1/N_c$ and can be neglected for large $N_c$. 
\end{widetext}


\begin{thebibliography}{9}
\bibitem{ritchie1957}
R. H. Ritchie, Phys. Rev. \textbf{106}, 874 (1957).
\bibitem{ferrell1958}
R. A. Ferrell, Phys. Rev. \textbf{111}, 1214 (1958).
\bibitem{raether1988}
H. Raether, \textit{Surface plasmons on smooth and rough surfaces and on gratings} (Springer Berlin Heidelberg,1988).
\bibitem{benno1988}
B. Rothenh\"{a}usler and K. Wolfgang, Nature \textbf{332}, 615 (1988).
\bibitem{zayats2005}
A. V. Zayats, I. S. Igor and A. A. Maradudin, Phys. Rep. \textbf{408}, 131 (2005).
\bibitem{maier2007}
S. A. Maier \textit{Plasmonics: fundamentals and applications} (Springer Science \& Business Media, 2007).
\bibitem{ozbay2006}
E. Ozbay, Science \textbf{311}: 189 (2006).
\bibitem{mark2007}
M. L. Brongersma and P. G. Kik, \textit{Surface plasmon nanophotonics} (Springer, 2007).
\bibitem{barnes2003}
W. L. Barnes, A. Dereux, and T. W. Ebbesen, Nature \textbf{424}, 824 (2003).
\bibitem{barnes2006}
W. L. Barnes, J. of Optics A \textbf{8}, S87 (2006).
\bibitem{ebbesen2008}
T. W. Ebbesen, C. Genet and S. I. Bozhevolnyi, Physics Today \textbf{61}, 44 (2008).
\bibitem{martino2012}
D. M. Giuliana, Y. Sonnefraud, S. K\'{e}na-Cohen, M. Tame, S. K. \"{O}zdmir, M. S. Kim and S. A. Maier, Nano Letters \textbf{12}, 2504 (2012).
\bibitem{bergman2003}
D. J. Bergman and M. I. Stockman, Phys. Rev. Lett. \textbf{90}, 027402 (2003).
\bibitem{seidel2005}
J. Seidel, S. Frafstron and L. Eng, Phys. Rev. Lett.\textbf{94}, 177401 (2005).
\bibitem{leon2008}
I. De Leon and P. Berini, Phys. Rev. B \textbf{78}, 161401(R) (2008).
\bibitem{leon2010}
I. De Leon and P. Berini, Nat. Photonics \textbf{4}, 382 (2010).
\bibitem{yu2011}
D. Y. Fedyanin and A. Y. Arsenin, Opt. Express \textbf{19}, 12524 (2011).
\bibitem{pierre2012}
P. Berini and I. De Leon, Nat. Photonics \textbf{6},16 (2012).
\bibitem{fedyanin2012}
D. Y. Fedyanin, A. V. Arsenin and A. V. Zayats, Nano Letters \textbf{12}, 2459 (2012).
\bibitem{cohen2013}
S. K\'{e}na-Cohen, P. N. Stavrinou, D. D. C. Bradley and S. A. Maier, Nano Letters \textbf{13}, 1323 (2013).
\bibitem{aryal2015}
C. M. Aryal, B. Y.-K. Hu and A.-P. Jauho, arXiv:1508.01271 (2015).
\bibitem{pitarke2007}
J. M. Pitarke, V. M. Silkin, E. V. Chulkov and P. M. Echenique, Rep. Prog. Phys. \textbf{70}, 1 (2007).
\bibitem{sarid2010}
D. Sarid and W. Cgallener, \textit{Modern introduction to surface plasmons: theory, mathematic modeling and applications} (Cambridge University Press, Cambridge, UK, 2010).
\bibitem{harris1971}
J. Harris, Phys. Rev. B \textbf{4}, 1022 (1971).
\bibitem{fetter}
A. L. Fetter, Ann. Physics \textbf{81}, 367 (1973); Phys. Rev. B \textbf{33}, 3717 (1986).
\bibitem{peter1984}
P. J. Feibelman, Prog. Surf. Science. \textbf{12}, 287 (1982).
\bibitem{schnitzer2016}O. Schnitzer, V. Giannini, S. A. Maier and R. V. Craster, Proc. R. Soc. A 472, 20160258 (2016). 
\bibitem{landau1946}
L. Landau, J. Phys. USSR \textbf{X}, 25 (1946).
\bibitem{wong1964}
A. Y. Wong, R. W. Motley and N. D'angelo, Phys. Rev. \textbf{133}, A436 (1964).
\bibitem{bingham1997}
R. Bingham and J. T. Mendonca and J. M. Dawson, Phys. Rev. Lett. \textbf{78}, 247 (1997).
\bibitem{natu2013}
S. S. Natu and R. M. Wilson, Phys. Rev. A \textbf{88}, 063638 (2013).
\bibitem{ziman1960}
J. M. Ziman, \textit{Electrons and Phonons: the theory of transport phenomena in solids} (Oxford University Press, 2001).
\bibitem{pines}
D. Pines, \textit{Elementary excitations in solids} (W. A. Benjamin, New York, 1963)
\bibitem{abrikosov}
A. A. Abrikosov, \textit{Fundamentals of the theory of metals} (Elsiver Science Publishers B. V., North-Holland, 1988)
\bibitem{reuter1948}
G. E. H. Reuter and E. H. Sondheimer, Proc. R. Soc. Lond. A \textbf{195}, 338 (1948).
\bibitem{kaganov1997}
M. I. Kaganov, G. Y. Lyubarskiy and A. G.  Mitina, Phys. Rep. \textbf{288}, 291 (1997).
\end{thebibliography}
\end{document}